\documentclass[]{aa}
\usepackage{graphicx}
\usepackage{natbib}

\usepackage{color}

\usepackage{amssymb}

\usepackage[varg]{txfonts}
\begin{document}

\definecolor{bleu}{rgb}{.3,0.2,1.0}
\definecolor{red}{rgb}{1.,0.2,0.2}

\title{A distinct magnetic property of the inner penumbral boundary}
\subtitle{Formation of a stable umbra-penumbra boundary in a sunspot}

\author{J. Jur\v{c}\'{a}k
        \inst{1}
        \and
        N. Bello Gonz\'{a}lez
        \inst{2}
        \and
        R. Schlichenmaier
        \inst{2}
        \and
        R. Rezaei
        \inst{2}}

\institute{Astronomical Institute of the Academy of Sciences, Fri\v{c}ova
  298, 25165 Ond\v{r}ejov, Czech Republic
  \and
  Kiepenheuer-Institut f\"{u}r Sonnenphysik, Sch\"{o}neckstr. 6, 79104 Freiburg, Germany}

\date{Received 11 December, 2014; accepted }

\abstract
    % context heading (optional); leave it empty if necessary
{A sunspot emanates from a growing pore or protospot. In order to trigger the formation of a penumbra, large inclinations at the outskirts of the protospot are necessary. The penumbra develops and establishes by colonising both umbral areas and granulation. Evidence for a unique stable boundary value for the vertical component of the magnetic field strength, $B^{\rm stable}_{\rm ver}$, was found along the umbra-penumbra boundary of developed sunspots.}
 {We study the changing value of $B_{\rm ver}$ as the penumbra forms and as it reaches a stable state. We compare this with the corresponding value in fully developed penumbrae.} 
 {We use broadband G-band images and spectropolarimetric GFPI/VTT data to study the evolution of and the vertical component of the magnetic field on a forming umbra-penumbra boundary. For comparison with stable sunspots, we also analyse the two maps observed by Hinode/SP on the same spot after the penumbra formed.}
{The vertical component of the magnetic field, $B_{\rm ver}$, at the umbra-penumbra boundary increases during penumbra formation owing to the incursion of the penumbra into umbral areas.  After 2.5 hours, the penumbra reaches a stable state as shown by the GFPI data. At this stable stage, the simultaneous Hinode/SP observations show a $B_{\rm ver}$ value comparable to that of umbra-penumbra boundaries of fully fledged sunspots.}
  {We confirm that the umbra-penumbra boundary, traditionally defined by an intensity threshold, is also characterised by a distinct canonical magnetic property, namely by  $B^{\rm stable}_{\rm ver}$. During the penumbra formation process, the inner penumbra extends into regions where the umbra previously prevailed. Hence, in areas where  $B_{\rm ver} < B^{\rm stable}_{\rm ver}$, the magneto-convection mode operating in the umbra turns into a penumbral mode. Eventually, the inner penumbra boundary settles at $B^{\rm stable}_{\rm ver}$, which hints toward the role of $B_{\rm ver}^{\rm stable}$ as inhibitor of the penumbral mode of magneto-convection.
}
  

\maketitle

%
%________________________________________________________________
%

\section{Introduction}
\label{introduction}

Since the discovery of the magnetic nature of sunspots \citep{Hale:1908, Hale:1909}, it became clear that the magnetic field strength decreases radially outward from the sunspot centre. The radial dependence  of the magnetic field strength and inclination (azimuthal averages around a spot) was investigated \citep[see review by][]{Solanki:2003} to estimate these parameters on the umbra-penumbra (UP) boundary. The reported values of the magnetic field strength and inclination on the UP boundary vary significantly \citep[e.g.][]{Lites:1990, Solanki:1992, Balthasar:1993, Keppens:1996, cwp:2001, Mathew:2003, Bellot:2003, Borrero:2004, Bellot:2004, Balthasar:2005, Sanchez:2005, Beck:2008}.

\citet{Jurcak:2011} finds a variation in the magnetic field strength and inclination along individual UP boundaries. However, the vertical component of the magnetic field ($B_{\rm ver}$) is found to be constant along all nine studied UP boundaries. The boundary values of $B_{\rm ver}$ change only slightly with sunspot size. There, Hinode/SP data were used which ensured that all nine sunspots were observed and analysed under identical conditions. We surmise that the finding of a constant value of $B_{\rm ver}$ along the UP boundary is only possible with a spectropolarimeter in space because varying conditions influence the results provided by ground-based instruments. 

The spectropolarimetric properties of a penumbra that formed in AR 11024 on July 4, 2009, as observed with GFPI/VTT were described in \citet{Rezaei:2012}. In \citet[][hereafter  Paper~I]{jurcak:2014a}, we found in the same  dataset that highly inclined magnetic fields trigger the formation of the penumbra. In this paper, we present the Hinode/SP measurements of this spot taken shortly after the penumbra was formed. We find that $B_{\rm ver}$ on the UP boundary is comparable to the value reported by \citet{Jurcak:2011}. For this reason, we extended the analysis of Paper~I by more than two hours to study how the growth of the penumbra proceeds and how $B_{\rm ver}$ saturates at a constant value.  We take advantage of the full GFPI/VTT dataset acquired between 08:32 UT and 12:40 UT, and compare it to the first two Hinode/SP maps of this sunspot that were taken at 12:30 UT and 15:44 UT. The data from VTT and Hinode/SP, and our analysis methods are described in Sect.~\ref{observations}. In Sect.~\ref{results} we describe the properties found on the forming UP boundary and  how $B_{\rm ver}$ saturates to reach the stable, {\em \emph{canonical}}, value. We discuss and summarise the results in Sect.~\ref{discussion}.

\begin{figure*}[!t]
 \centering \includegraphics[width=0.95\linewidth]{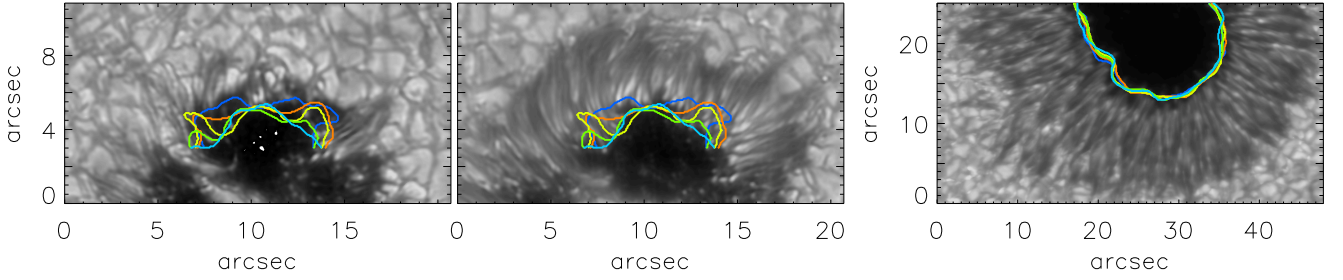}
 \caption{G-band images showing forming penumbra in active region NOAA~11024 at 8:35~UT (left panel) and 11:50~UT (middle panel) and developed penumbra in active region NOAA~10940 (right panel). The colour contours mark the position of the UP boundary at different times, and are marked by lines of respective colours in Fig.~\ref{boundary_drift}.}
 \label{G-band}
\end{figure*}

\section{Observations and data analysis}
\label{observations}

Our analysis is based on spectropolarimetric data taken with the GFPI at the German VTT \citep{Puschmann:2006, Nazaret:2008} along with G-band images. With the GFPI, we observed the four Stokes profiles of the Fe~I~617.3~nm line along 31 wavelength points. The resulting S/N ratio is 125 in the Stokes Q, U, and V profiles and the spatial resolution is around 0\farcs4. Details on the data reduction can be found in \citet{Rezaei:2012}. The G-band images were speckle reconstructed using the code KISIP \citep{Woger:2008}, have a cadence of 17~sec, and the resulting spatial resolution is better than 0\farcs3 \citep[see][]{Schlichenmaier:2010}. The forming sunspot was observed  on July 4, 2009, in the active region NOAA~11024, located at 6$^\circ$~E and 25$^\circ$~S from disc centre.

In Paper I, we studied 2\,h of penumbra formation, from 08:32\,UT to 10:20\,UT. In this work we extend our analysis to the full dataset, until 12:40\,UT. We also make use of two datasets of the same active region acquired by Hinode/SP \citep{Kosugi:2007, Tsuneta:2008} at 12:30~UT and 15:44~UT. The Hinode/SP recorded the Stokes profiles of the two \ion{Fe}{I} lines at 630.15 and 630.25~nm with a pixel sampling of 0\farcs32 and a noise level of $10^{-3}I_{\rm c}$. To illustrate the apparent motions in the developed sunspot and for comparison purposes, we use the Hinode G-band images of a sunspot in the active region NOAA~10940 observed on February 2, 2007. The images have a spatial resolution of 0\farcs22, and a cadence of 90~sec. The data were calibrated with the standard routines available in the SolarSoft Hinode package. 
%We use the Hinode datasets for comparison and completeness purposes. 

The spectropolarimetric data were inverted using the VFISV code \citep{Borrero:2011a}. It is a Milne-Eddington code, i.e. plasma parameters are constant with height and the source function is linearly dependent on optical depth. The same inversion scheme was used to invert both the GFPI and Hinode data. The magnetic filling factor is set to unity for all inverted pixels and we do not take into account a stray-light component.  

We used local correlation tracking \citep[LCT,][]{November:1988} to study horizontal motions in the G-band images. We first aligned both the VTT and Hinode G-band images and removed the p-mode oscillations by applying a $k-\omega$~filter with a cut-off of 5~km~s$^{-1}$. We used a Gaussian tracking window of FWHM 0\farcs7, which is fine enough to obtain the proper motions in penumbra.

\section{Results}
\label{results}

\subsection{The forming UP boundary vs. the stable UP boundary}
\label{migration}

In Fig.~\ref{G-band} we show the evolution of the UP boundary position in the forming penumbra in AR\,11024 (colour contours in left and middle panel) and compare it with the evolution in a developed sunspot (AR\,10940, right panel). The contours mark the position of the intensity boundaries at 50\% of the mean quiet Sun intensity of the spatially smoothed G-band images. Comparing the left and middle panel, one can see that the penumbra forms mostly at the expense of granular regions, but the changing contours also depict an umbral area that transformed into a penumbra (Paper~I). In the case of developed penumbra (right panel in Fig.~\ref{G-band}), there is no apparent motion of the boundary.

\begin{figure}[!t]
 \centering \includegraphics[width=0.9\linewidth]{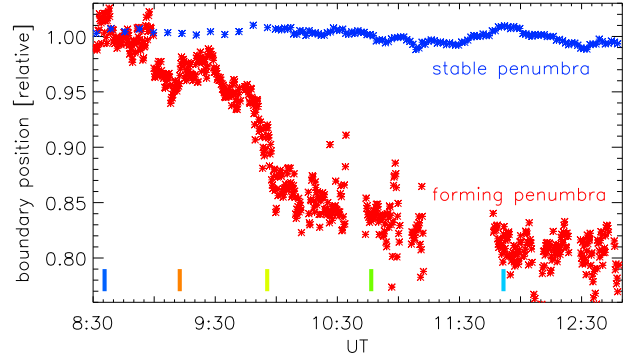}
 \caption{Evolution of the mean distance between the UP boundary and the sunspot centre (normalised to this distance). The red and blue symbols correspond to the forming and stable penumbra, respectively. The coloured vertical lines mark the times at which the contours in Fig.~\ref{G-band} are displayed. The UT time on the $x$-axis corresponds to the observations of the forming penumbra on July 4, 2009.} 
 \label{boundary_drift}
\end{figure}

In Fig.~\ref{boundary_drift}, we show the evolution of the mean distance of the UP boundary to the sunspot centre. In the developed penumbra, the UP boundary is stationary and we observe fluctuations of its position within $\pm1\%$ of the mean distance to the sunspot centre. In the forming penumbra, the mean distance between the UP boundary and the sunspot centre shows a decrease of approximately 20$\%$ within 3.5~hours. In the first hour, there are no stable penumbral segments formed and also no significant changes of the boundary position, as also found in Paper~I. Between 9:30~UT and 11:00~UT, we observe a rapid formation of penumbral filaments in the upper left and right regions of umbral core (compare areas around [8\arcsec, 5\arcsec] and [13\arcsec, 5\arcsec] in left and middle panel of Fig.~\ref{G-band}) that results in a decrease in the distance between the UP boundary and the sunspot centre (see also Paper~I). In the last segment of the G-band images (after 11:45~UT), the UP boundary reaches a stationary state. 

\begin{figure}[!t]
 \centering \includegraphics[width=0.85\linewidth]{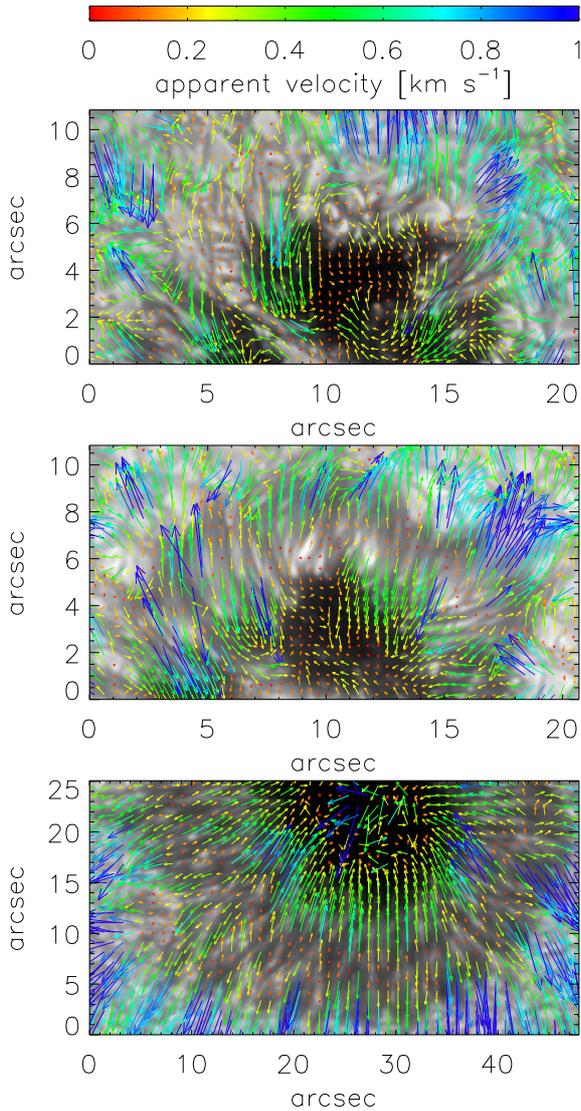}
 \caption{Apparent motions of brightness structures identified by the LCT method. The upper panel shows the apparent motions at the beginning of the penumbra formation (8:30--9:00~UT), the middle panel around the end of the penumbra formation (11:45--12:15~UT), and the bottom panel shows the apparent motions in developed penumbra for comparison.}
 \label{lct}
\end{figure}

As shown in Fig.~\ref{lct}, we observe apparent motions of penumbral grains toward the umbra both in forming and stable penumbrae. This behaviour is well known for developed penumbrae and the typical amplitude of these motions is around 0.4~km~s$^{-1}$, which is in agreement with previous studies \citep{Wang:1992, Sobotka:1999, MArquez:2006, jurcak:2014}. 

In sectors where the penumbra has already formed, we find inward motions in the inner penumbra and outward motions in the outer penumbra (see  [16\arcsec, 3\arcsec] in upper plot of Fig.~\ref{lct}), as in developed sunspots (bottom panel of Fig.~\ref{lct}). 
In sectors where penumbral filaments form between 8:30~UT and 9:00~UT at the left side of the umbral core [8\arcsec, 5\arcsec], we find inward motions with amplitudes around 0.4~km~s$^{-1}$. These are not yet accompanied by outward motions. At the upper part of the umbral core [10\arcsec-15\arcsec, 6\arcsec], there are no stable penumbral filaments observed before 9:30~UT and there we find slower and not organised apparent motions. In the umbra, the arrows display a swirl motion of umbral dots. The latter is investigated in \citet{Nazaret:2012}.

In the middle plot of Fig.~\ref{lct}, we show apparent motions around 12:00~UT, when the whole segment of the penumbra was developed. We find inward motions in the inner penumbra and outward motions in the outer penumbra as in developed sunspots. The only exception is the region around [9\arcsec, 6\arcsec], where the penumbral grains do not show any motions toward the umbra. After investigating the sequence of G-band images, it appears that the apparent motions in this region are blocked by the evolution of penumbral grains in curved filaments at [8\arcsec, 4\arcsec], which protrude into the same region. 

In developed sunspots, the penumbral grains disappear around the UP boundary \citep[or transform into umbral dots,][]{Sobotka:2009} and the boundary position does not change over time. However, the apparent motions of the penumbral grains in the forming penumbra do not cease at the UP boundary, but result in the migration of this boundary toward the sunspot centre. After 12:00~UT the UP boundary reaches a stable position, which is no longer influenced by the continuous inward motions of penumbral grains. This finding can be naturally explained in the context of the rising of the moving tube by \cite{Schlichenmaier:1998} (see Sect.\,\ref{discussion}).

\subsection{The vertical component of the magnetic field in the UP boundary}
\label{magnetic_field}

\begin{figure}[!t]
 \centering \includegraphics[width=0.91\linewidth]{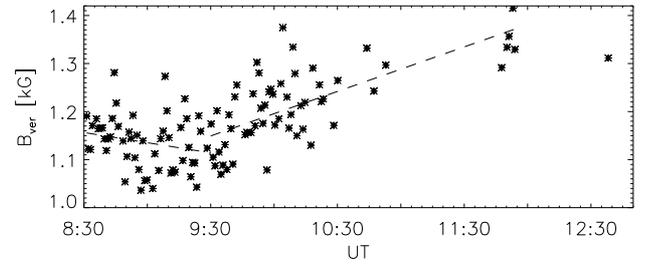}
 \caption{Temporal evolution of $B_{\rm ver}$. Symbols correspond to mean values of $B_{\rm ver}$ along the UP boundary for each GFPI scan, where examples of the UP boundaries are marked by colour contours in Fig.~\ref{G-band}. Dashed lines correspond to the linear fits of the obtained values between 8:30~UT and 9:30~UT (before the rapid penumbra evolution) and between 9:30~UT and 12:00~UT.}
 \label{b_v}
\end{figure}

According to \citet{Jurcak:2011}, the stable value of the vertical component of the magnetic field strength ($B_{\rm ver}^{\rm stable}$) defines the UP boundary in developed sunspots. Here, we investigate the evolution of $B_{\rm ver}$ for an extended time interval of 4.5\,hours, during which the penumbra reaches a steady state (see Fig.\,\ref{b_v}).

Before 9:30~UT, no stable penumbral filaments have formed yet, and neither the boundary position (Fig.~\ref{boundary_drift}) nor the mean $B_{\rm ver}$  changes significantly. Around 9:30~UT, we observe the weakest $B_{\rm ver}$ of about 1100~G. As the UP boundary migrates toward the sunspot centre, $B_{\rm ver}$ increases to approximately 1340~G at 11:50~UT. Fifty minutes later, at the end of the observations, $B_{\rm ver}$ shows a similar value of 1310~G. This, along with the stationary position of the UP boundary after 12:00~UT (see Fig.~\ref{boundary_drift}), suggests that the forming UP boundary has reached a stable state. 

At 12:30, the same sunspot in AR\,11024 was observed with Hinode/SP. After a careful alignment of the datasets, we find an average value of $B_{\rm ver}$ of 1810~G on the same segment of the UP boundary. Moreover, a Hinode/SP scan taken later at 15:45~UT shows a comparable value of $B_{\rm ver}$ of 1830~G. These values are in agreement with the reported values of $B_{\rm ver}^{\rm stable}$ on the UP boundaries of small sunspots \citep[cf. Table 5 in][]{Jurcak:2011}. Hence, the stability of the $B_{\rm ver}$ value observed by the GFPI/VTT is confirmed by the Hinode/SP results.

The stable value of $B_{\rm ver}$ on the UP boundary measured by the GFPI/VTT is approximately 500\,G smaller than that measured by Hinode on the same spot. The difference in $B$ between GFPI and Hinode/SP is around 150~G. This can be ascribed to systematic effects, e.g. different instrument characteristics, spectral line, and stray light. The larger difference we find for $B_{\rm ver}$ is accompanied by a corresponding difference in inclination. We speculate that this difference might be due to limitations in the polarimetric calibration of the GFPI data. However, this fact does not understate the relevance of the main findings, i.e. (1) the increase of $B_{\rm ver}$ on the UP boundary during penumbra formation and (2) the stable state reached by this boundary at the distinctive value $B_{\rm ver}^{\rm stable} $, common to all sunspots.

\section{Discussion}
\label{discussion}

\subsection{A canonical value for $B_{\rm ver}$ at the inner penumbral end}

Analysing the UP boundary in nine stable sunspots, \citet{Jurcak:2011} finds evidence that $B_{\rm ver}$ on this boundary is of similar magnitude in all sunspots of his sample. Using Hinode/SP data, he finds a value of $B_{\rm ver}^{\rm stable}=$1860 ($\pm 190$) G. This constancy is  remarkable. It implies that the UP boundary, which is traditionally defined by an intensity threshold, is also characterised by a distinct magnetic property, $B_{\rm ver}^{\rm stable}$.

 We confirm this result. Studying the behaviour of the vertical component of the magnetic field in the UP boundary during the process of penumbra formation, we find that
(1) once the onset of the penumbra takes place, the forming UP boundary migrates toward the umbra and the corresponding $B_{\rm ver}$ increases accordingly; 
(2) after 2.5\,h, the migration ceases at a certain $B_{\rm ver}$ value. From then on, the UP boundary position and $B_{\rm ver}$ are preserved; and  
(3) Hinode/SP observations of the same sunspot show a stationary value of $B_{\rm ver}$ at the same boundary. This Hinode value is in agreement with those found on UP boundaries of developed sunspots \citep[][also using Hinode]{Jurcak:2011}. Hence, we conclude that the UP boundary has reached a stable state characterised by $B_{\rm ver}^{\rm stable}$.

In order to trigger the formation of penumbral filaments, large inclinations of the magnetic field are involved \citep[][Paper I]{Rezaei:2012}. It is observed that penumbral filaments establish by expanding into granular and umbral areas. In the inner part, the penumbral mode of magneto-convection extends into regions where previously the umbral mode of magneto-convection prevailed, but only until a critical boundary value, $B_{\rm ver}^{\rm stable}$, is reached. 

Hence, during the formation of the penumbra, the protospot is partially converted into penumbra. We propose the following rule: 
\begin{enumerate}
\item The penumbral mode of magneto-convection takes over in areas with $B_{\rm ver} < B_{\rm ver}^{\rm stable}$.
\item The umbral mode of magneto-convection prevails in areas with $B_{\rm ver} > B_{\rm ver}^{\rm stable}$.
\end{enumerate}

Since $B_{\rm ver}^{\rm stable}$ appears to be common to all stable sunspots, we can define $B_{\rm ver}^{\rm stable}$ as a {\em \emph{canonical}} value. It differs depending on systematic differences that arise from the observational setup, the spectral line, the seeing, calibration issues, and from the method of analysis. Therefore, Hinode/SP is an ideal tool for inferring a reference value of $B^{\rm stable}_{\rm ver}$ as it avoids most of these problems. 

\subsection{Boundary and penumbral filaments}

In a forming penumbral segment, filaments grow and expand. From their first appearance, penumbral grains exist at their inner end. Their inward motion leads the expansion of the forming filaments into the umbral areas. This  may provide an explanation to the question of how the inner end of a penumbra is determined. In the framework of the {\it moving tube model} \citep{Schlichenmaier:1998, Schlichenmaier:2002}, once a critical inclination of the magnetic field at the outer edge of the protospot is reached, individual flux tubes become buoyant and rise.  As the tubes rise, their cross sections with the photosphere (penumbral grains) migrate inward to areas that were previously stable and develop an up- and outflow along the tube. As the subphotospheric part of the tube becomes more and more vertical, buoyancy diminishes and the rising phase as well as the inward migration terminates. Hence, once the penumbral mode of convection starts, the penumbra grows inward at the cost of umbral area.

\subsection{Concluding remark}
In this paper we find further evidence for the existence of a canonical $B_{\rm ver}^{\rm stable}$ characterising the UP boundary. This discovery has important implications for our understanding of magneto-convection in sunspots as it hints toward $B_{\rm ver}^{\rm stable}$ playing a crucial role as inhibitor of the penumbral mode of magneto-convection. Therefore, to improve the statistics  it is essential to study a larger sample of sunspots and to make a detailed analysis of the coupling of the thermodynamical and magnetic properties at the UP boundary. 

\begin{acknowledgements}

We are grateful to Juan Manuel Borrero for providing us with the inversion results obtained by his code VFISV. The support from GA~CR~P209/12/0287, GA~CR~14-04338S, and RVO:67985815 is gratefully acknowledged. N.B.G. acknowledges financial support by the Senatsausschuss of the Leibniz-Gemeinschaft, Ref.-No. SAW-2012-KIS-5. R.R. acknowledges financial support by the DFG grant RE 3282/1-1. The Vacuum Tower Telescope is operated by the Kiepenheuer-Institut f\"{u}r Sonnenphysik, Freiburg, at the Spanish Observatorio del Teide of the Instituto de Astrof\'{\i}sica de Canarias. Hinode is a Japanese mission developed and launched by ISAS/JAXA, with NAOJ as domestic partner and NASA and STFC (UK) as international partners. It is operated by these agencies in cooperation with ESA and NSC (Norway).

\end{acknowledgements}

\bibliographystyle{aa}
\bibliography{25501}

\begin{thebibliography}{34}
\expandafter\ifx\csname natexlab\endcsname\relax\def\natexlab#1{#1}\fi

\bibitem[{{Balthasar} \& {Collados}(2005)}]{Balthasar:2005}
{Balthasar}, H. \& {Collados}, M. 2005, \aap, 429, 705

\bibitem[{{Balthasar} \& {Schmidt}(1993)}]{Balthasar:1993}
{Balthasar}, H. \& {Schmidt}, W. 1993, \aap, 279, 243

\bibitem[{{Beck}(2008)}]{Beck:2008}
{Beck}, C. 2008, \aap, 480, 825

\bibitem[{{Bello Gonz{\'a}lez} \& {Kneer}(2008)}]{Nazaret:2008}
{Bello Gonz{\'a}lez}, N. \& {Kneer}, F. 2008, \aap, 480, 265

\bibitem[{{Bello Gonz{\'a}lez} {et~al.}(2012){Bello Gonz{\'a}lez}, {Kneer}, \&
  {Schlichenmaier}}]{Nazaret:2012}
{Bello Gonz{\'a}lez}, N., {Kneer}, F., \& {Schlichenmaier}, R. 2012, \aap, 538,
  A62

\bibitem[{{Bellot Rubio} {et~al.}(2004){Bellot Rubio}, {Balthasar}, \&
  {Collados}}]{Bellot:2004}
{Bellot Rubio}, L.~R., {Balthasar}, H., \& {Collados}, M. 2004, A\&A, 427, 319

\bibitem[{{Bellot Rubio} {et~al.}(2003){Bellot Rubio}, {Balthasar}, {Collados},
  \& {Schlichenmaier}}]{Bellot:2003}
{Bellot Rubio}, L.~R., {Balthasar}, H., {Collados}, M., \& {Schlichenmaier}, R.
  2003, \aap, 403, L47

\bibitem[{{Borrero} {et~al.}(2004){Borrero}, {Solanki}, {Bellot Rubio}, {Lagg},
  \& {Mathew}}]{Borrero:2004}
{Borrero}, J.~M., {Solanki}, S.~K., {Bellot Rubio}, L.~R., {Lagg}, A., \&
  {Mathew}, S.~K. 2004, A\&A, 422, 1093

\bibitem[{{Borrero} {et~al.}(2011){Borrero}, {Tomczyk}, {Kubo},
  {Socas-Navarro}, {Schou}, {Couvidat}, \& {Bogart}}]{Borrero:2011a}
{Borrero}, J.~M., {Tomczyk}, S., {Kubo}, M., {et~al.} 2011, \solphys, 273, 267

\bibitem[{{Hale}(1908)}]{Hale:1908}
{Hale}, G.~E. 1908, ApJ, 28, 315

\bibitem[{{Hale}(1909)}]{Hale:1909}
{Hale}, G.~E. 1909, \pasp, 21, 205

\bibitem[{{Jur{\v c}{\'a}k}(2011)}]{Jurcak:2011}
{Jur{\v c}{\'a}k}, J. 2011, \aap, 531, A118

\bibitem[{{Jur{\v c}{\'a}k} {et~al.}(2014{\natexlab{a}}){Jur{\v c}{\'a}k},
  {Bello Gonz{\'a}lez}, {Schlichenmaier}, \& {Rezaei}}]{jurcak:2014a}
{Jur{\v c}{\'a}k}, J., {Bello Gonz{\'a}lez}, N., {Schlichenmaier}, R., \&
  {Rezaei}, R. 2014{\natexlab{a}}, \pasj

\bibitem[{{Jur{\v c}{\'a}k} {et~al.}(2014{\natexlab{b}}){Jur{\v c}{\'a}k},
  {Bellot Rubio}, \& {Sobotka}}]{jurcak:2014}
{Jur{\v c}{\'a}k}, J., {Bellot Rubio}, L.~R., \& {Sobotka}, M.
  2014{\natexlab{b}}, \aap, 564, A91

\bibitem[{{Keppens} \& {Martinez Pillet}(1996)}]{Keppens:1996}
{Keppens}, R. \& {Martinez Pillet}, V. 1996, \aap, 316, 229

\bibitem[{{Kosugi} {et~al.}(2007){Kosugi}, {Matsuzaki}, {Sakao}, {Shimizu},
  {Sone}, {Tachikawa}, {Hashimoto}, {Minesugi}, {Ohnishi}, {Yamada}, {Tsuneta},
  {Hara}, {Ichimoto}, {Suematsu}, {Shimojo}, {Watanabe}, {Shimada}, {Davis},
  {Hill}, {Owens}, {Title}, {Culhane}, {Harra}, {Doschek}, \&
  {Golub}}]{Kosugi:2007}
{Kosugi}, T., {Matsuzaki}, K., {Sakao}, T., {et~al.} 2007, \solphys, 243, 3

\bibitem[{{Lites} {et~al.}(1990){Lites}, {Skumanich}, \&
  {Scharmer}}]{Lites:1990}
{Lites}, B.~W., {Skumanich}, A., \& {Scharmer}, G.~B. 1990, ApJ, 355, 329

\bibitem[{{M{\'a}rquez} {et~al.}(2006){M{\'a}rquez}, {S{\'a}nchez Almeida}, \&
  {Bonet}}]{MArquez:2006}
{M{\'a}rquez}, I., {S{\'a}nchez Almeida}, J., \& {Bonet}, J.~A. 2006, \apj,
  638, 553

\bibitem[{{Mathew} {et~al.}(2003){Mathew}, {Lagg}, {Solanki}, {Collados},
  {Borrero}, {Berdyugina}, {Krupp}, {Woch}, \& {Frutiger}}]{Mathew:2003}
{Mathew}, S.~K., {Lagg}, A., {Solanki}, S.~K., {et~al.} 2003, A\&A, 410, 695

\bibitem[{{November} \& {Simon}(1988)}]{November:1988}
{November}, L.~J. \& {Simon}, G.~W. 1988, \apj, 333, 427

\bibitem[{{Puschmann} {et~al.}(2006){Puschmann}, {Kneer}, {Seelemann}, \&
  {Wittmann}}]{Puschmann:2006}
{Puschmann}, K.~G., {Kneer}, F., {Seelemann}, T., \& {Wittmann}, A.~D. 2006,
  \aap, 451, 1151

\bibitem[{{Rezaei} {et~al.}(2012){Rezaei}, {Bello Gonz{\'a}lez}, \&
  {Schlichenmaier}}]{Rezaei:2012}
{Rezaei}, R., {Bello Gonz{\'a}lez}, N., \& {Schlichenmaier}, R. 2012, \aap,
  537, A19

\bibitem[{{S{\'a}nchez Cuberes} {et~al.}(2005){S{\'a}nchez Cuberes},
  {Puschmann}, \& {Wiehr}}]{Sanchez:2005}
{S{\'a}nchez Cuberes}, M., {Puschmann}, K.~G., \& {Wiehr}, E. 2005, A\&A, 440,
  345

\bibitem[{{Schlichenmaier}(2002)}]{Schlichenmaier:2002}
{Schlichenmaier}, R. 2002, Astronomische Nachrichten, 323, 303

\bibitem[{{Schlichenmaier} {et~al.}(1998){Schlichenmaier}, {Jahn}, \&
  {Schmidt}}]{Schlichenmaier:1998}
{Schlichenmaier}, R., {Jahn}, K., \& {Schmidt}, H.~U. 1998, A\&A, 337, 897

\bibitem[{{Schlichenmaier} {et~al.}(2010){Schlichenmaier}, {Rezaei}, {Bello
  Gonz{\'a}lez}, \& {Waldmann}}]{Schlichenmaier:2010}
{Schlichenmaier}, R., {Rezaei}, R., {Bello Gonz{\'a}lez}, N., \& {Waldmann},
  T.~A. 2010, \aap, 512, L1+

\bibitem[{{Sobotka} {et~al.}(1999){Sobotka}, {Brandt}, \&
  {Simon}}]{Sobotka:1999}
{Sobotka}, M., {Brandt}, P.~N., \& {Simon}, G.~W. 1999, A\&A, 348, 621

\bibitem[{{Sobotka} \& {Jur{\v c}{\'a}k}(2009)}]{Sobotka:2009}
{Sobotka}, M. \& {Jur{\v c}{\'a}k}, J. 2009, \apj, 694, 1080

\bibitem[{{Solanki}(2003)}]{Solanki:2003}
{Solanki}, S.~K. 2003, A\&A Rev., 11, 153

\bibitem[{{Solanki} {et~al.}(1992){Solanki}, {R\"{u}edi}, \&
  {Livingston}}]{Solanki:1992}
{Solanki}, S.~K., {R\"{u}edi}, I., \& {Livingston}, W. 1992, A\&A, 263, 339

\bibitem[{{Tsuneta} {et~al.}(2008){Tsuneta}, {Ichimoto}, {Katsukawa}, {Nagata},
  {Otsubo}, {Shimizu}, {Suematsu}, {Nakagiri}, {Noguchi}, {Tarbell}, {Title},
  {Shine}, {Rosenberg}, {Hoffmann}, {Jurcevich}, {Kushner}, {Levay}, {Lites},
  {Elmore}, {Matsushita}, {Kawaguchi}, {Saito}, {Mikami}, {Hill}, \&
  {Owens}}]{Tsuneta:2008}
{Tsuneta}, S., {Ichimoto}, K., {Katsukawa}, Y., {et~al.} 2008, \solphys, 249,
  167

\bibitem[{{Wang} \& {Zirin}(1992)}]{Wang:1992}
{Wang}, H. \& {Zirin}, H. 1992, \solphys, 140, 41

\bibitem[{{Westendorp Plaza} {et~al.}(2001){Westendorp Plaza}, {del Toro
  Iniesta}, {Ruiz Cobo}, {Pillet}, {Lites}, \& {Skumanich}}]{cwp:2001}
{Westendorp Plaza}, C., {del Toro Iniesta}, J.~C., {Ruiz Cobo}, B., {et~al.}
  2001, ApJ, 547, 1130

\bibitem[{{W{\"o}ger} \& {von der L{\"u}he}(2008)}]{Woger:2008}
{W{\"o}ger}, F. \& {von der L{\"u}he}, II, O. 2008, in Society of Photo-Optical
  Instrumentation Engineers (SPIE) Conference Series, Vol. 7019, Society of
  Photo-Optical Instrumentation Engineers (SPIE) Conference Series

\end{thebibliography}

\end{document}